\documentclass[journal=jacsat,manuscript=article]{achemso}

\usepackage[version=3]{mhchem} 
\usepackage{booktabs}
\usepackage{cleveref}
\usepackage{graphicx}
\usepackage[T1]{fontenc}
\usepackage[utf8]{inputenc}
\usepackage{mathtools}
\usepackage{multirow}
\usepackage{tabularx}
\usepackage[normalem]{ulem}
\usepackage{amssymb}
\usepackage{bm}
\usepackage{siunitx}
\usepackage{multirow}
\usepackage{xcolor}

\def\etal{{\em et al.\/}}
\def\crsbr{CrSBr}
\def\btef{BTE$_\text{full}$}

\author{Marta Loletti}
\affiliation[ICMAB]{Institut de Ciència de Materials de Barcelona, 
             ICMAB-CSIC, Campus UAB, 08193 Bellaterra, Spain}

\author{Alejandro Molina-S\'anchez}
\affiliation[UV]{Institute of Materials Science (ICMUV), University of 
                 Valencia, Catedr\'atico Beltr\'an 2, E-46980 Valencia, Spain}

\author{Juan Sebasti\'an Reparaz}
\affiliation[ICMAB]{Institut de Ciència de Materials de Barcelona, 
                    ICMAB-CSIC, Campus UAB, 08193 Bellaterra, Spain}

\author{Xavier Cartoix\`a}
\affiliation[UAB]{Departament d’Enginyeria Electr\`onica, Universitat 
                  Aut\`onoma de Barcelona, Bellaterra, 08193, Barcelona, Spain}

\author{Riccardo Rurali}
\affiliation[ICMAB]{Institut de Ciència de Materials de Barcelona, 
                    ICMAB-CSIC, Campus UAB, 08193 Bellaterra, Spain}
\email{rrurali@icmab.es}

\title{Thermal conductivity and tunable thermal anisotropy of magnetic CrSBr monolayer}


\begin{document}

%

%
%
%

\begin{abstract}
We present first-principles calculations of the thermal conductivity,
${\bm \kappa}$, of monolayer CrSBr, a van der Waals magnetic 2D material.
We find a considerable thermal anisotropy, with a ratio $\kappa_{xx}/\kappa_{yy}$
of around 1.8. The anisotropy stems from a combined effect of phonon
velocities and lifetimes and can be tuned by controlling the flake size
by suppressing long mean path phonons.
\end{abstract}

\section{\label{sec:intro}Introduction}

The advent of two-dimensional materials (2DMs)~\cite{NovoselovJiangSchedin2005, 
NovoselovScience16, Novoselov2011} has opened up many exciting possibilities, 
including channels for field effect transistors with improved gate 
control~\cite{RadisavljevicRadenovicBrivio2011}, van der Waals 
heterostructures~\cite{GeimGrigorieva2013} for efficient photodetectors~\cite{NovoselovScience16}, 
and non-volatile spintronic memories~\cite{Yang2022}. With the progressive prediction 
and synthesis of 2DMs featuring different possible electronic properties 
(i.e., metallic, semiconductor, topological insulators) as long-established 
3D material counterparts, it was only a matter of time until the first 
single-/few-layer magnetic 2DMs ~\cite{Gibertini2019} were experimentally 
found, namely CrGeTe$_3$~\cite{Gong2017} and CrI$_3$~\cite{Huang2017}. 
Since then, many others have been added to the catalog of 2D 
magnets~\cite{WangQingBedoya2022} and they are now the focus of an 
intense research activity. A question that naturally arises is how the 
magnetic order couples to other physical properties. Telford 
\etal~\cite{Telford2022} have found that the magnetoresistance presents 
clear signatures of the underlying magnetic order, whereas Peng 
\etal~\cite{PengYuxuan2022} showed that the spin orientation and 
the metamagnetic transitions in CrPS$_4$ can be controlled by 
hydrostatic pressure. The interplay of magnetism and thermal transport 
properties in 2D magnets, however, is an area that has received little 
attention so far. 

Among van der Waals magnetic materials, CrSBr stands out due to robust 
magnetism, air-stability, strong anisotropy, and quasi one-dimensional 
electron transport~\cite{ZiebelNL24}. Similar to other 2D magnetic
materials (e.g. CrI$_3$), in CrSBr each layer is ordered ferromagnetically
and couples antiferromagnetically with the other layers~\cite{LeeNL21}. 
Due to the interest sparked by these properties, very recently CrSBr 
thermal transport properties have been the focus of a few theoretical
studies, but the results feature a huge variability. Xuan \etal~\cite{XuanMA23}
reported an ultralow lattice thermal conductivity, ${\bm \kappa}$, for 
CrSX (X = Cl, Br, I) monolayers (MLs), as obtained from density-functional
theory calculations (DFT) and the solution of the Boltzmann Transport 
Equation (BTE). Although they reported values of ${\bm \kappa}$ {\it above} 
the Curie temperature of ML CrSBr (experimentally estimated to be 146~K by 
Lee \etal~\cite{LeeNL21}), by roughly extrapolating their results at 200~K 
down to lower temperatures, one obtains $\kappa_{xx} \approx 
0.8$~W~m$^{-1}$K$^{-1}$  and $\kappa_{yy} \approx 0.3$~W~m$^{-1}$K$^{-1}$.
More recently, however, Liu \etal~\cite{LiuJPCM25} reported significantly larger
values, 25.48 and 11.90~W~m$^{-1}$K$^{-1}$ at 150~K for $\kappa_{xx}$ and
$\kappa_{yy}$, respectively. This difference is striking, because both
groups use the same DFT code and exchange-correlation functional. The
linearized BTE, on the other hand, is solved with \texttt{ShengBTE}~\cite{LiCPC14} 
in one case and phono3py~\cite{TogoJPCM23} in the other, but a recent 
comparative study allows ruling out that this different choice could 
originate sizeable discrepancies~\cite{McGaugheyJAP25}.
Larger thermal conductivities --yet of the same order of magnitude
as those of Ref.~\citenum{LiuJPCM25}-- have been obtained by Han and
coworkers~\cite{HanJAP25}, who report values of 80.79 and 
37.38~W~m$^{-1}$K$^{-1}$ at 145~K. In this work, a machine learning
force-field is trained on DFT datasets and then used to
compute harmonic and anharmonic force constants. Noteworthy, this study
included both three- and four-phonon scattering, thus accounting
for one additional scattering mechanism. Yet, they found  
larger ${\bm \kappa}$ than Liu and coworkers~\cite{LiuJPCM25}, who only
considered anharmonicity up to the third-order. 




Here we revisit the calculation of the thermal conductivity 
of ML \crsbr, focusing our attention on the anisotropy of heat transport.
We first study the robustness of the ferromagnetic (FM) ground-state 
and explore the possibility of observing a strain-induced phase 
transition, where the antiferromagnetic (AFM) ordering can become 
favored, considering both uniaxial and biaxial strain conditions. 
This would be an important feature, because one could in principle 
selectively access both magnetic phases by triggering an AFM~$\rightarrow$~FM 
transition with an external magnetic field. We then present a detailed analysis 
of the thermal conductivity of ML CrSBr, studying both the dependence 
on temperature and on the flake size and shedding light on the reason of 
the anisotropy of the in-plane thermal conductivity. Finally, we highlight 
how dimensional confinement can be used as a control knob to tune the 
anisotropy of heat transport. 


\section{\label{sec:methods}Computational Methods}

Calculations have been carried out within density functional theory (DFT) \cite{HohenbergKohn1964,KohnSham1965} as implemented in the Vienna \textit{ab initio} simulation package \cite{VASPKIT}, with the projector augmented wave (PAW) pseudopotential method and a plane-wave cutoff of 520 eV. The exchange-correlation energy was described by the Perdew-Burke-Ernzerhof (PBE) ~\cite{PerdewBurkeErnzerhof1996} functional within the framework of the generalized gradient approximation (GGA), and the long-range dispersion interactions were included through van der Waals corrections described in the framework of Grimme's approach with zero-damping function (DFT-D3)\cite{GrimmeStefanAntony}.
To properly describe the localized $d$-electrons of the transition-metal atoms, we employed the DFT+U approach as implemented in the VASP code. The on-site Coulomb interaction was treated using the simplified rotationally invariant formalism of Dudarev {\it et al.}~\cite{Dudarev_at_all} and the correction was applied with an effective Hubbard parameter $U_{eff}$ = 4.0 eV~\cite{YangPRB21, LiuJPCM25}.
The monolayer was modeled using approximately 20 \AA\space vacuum along the out-of-plane, {\bf c}-direction to avoid spurious interactions between periodic images. A 12$\times$10$\times$1 Monkhorst-Pack ${\bf k}$-grid was used for Brillouin zone integration. Convergence was reached when forces were lower than 0.005 eV/\AA\ and stress lower than 0.03~kbar.
Second- and third-order interatomic force constants--- IFC2s and IFC3s, respectively---were used as input for the almaBTE suite ~\cite{CarreteVermeerschKatre2017} to iteratively solve the Boltzmann Transport Equation (BTE) beyond the relaxation time approximation (RTA). 
IFC2s and the subsequent phonon dispersion curves were computed by finite differences with the \texttt{phonopy} software~\cite{TogoJPCM23, phonopy2-phono3py-JPSJ}, using a 6$\times$5$\times$1 supercell. IFC3s were also computed by finite differences using the same supercell,
with the \texttt{thirdorder.py} package~\cite{LiCPC14}. IFC2s were postprocessed with the hiPhive package~\cite{ErikssonATS19} to enforce the constraints imposed by rotational sum rules, necessary to obtain the correct quadratic dispersion for the ZA modes in 2DMs~\cite{CarreteMRL16}.
To reproduce the suppression of LO-TO splitting at the $\Gamma$-point in 2D materials~\cite{SanchezPortalPRB02, SohierNL17, DeLuca2DMater20}, phonons dispersions are computed without including non-analytical corrections in their standard formalism.

\begin{figure}[t]
\includegraphics[width=0.65\linewidth]{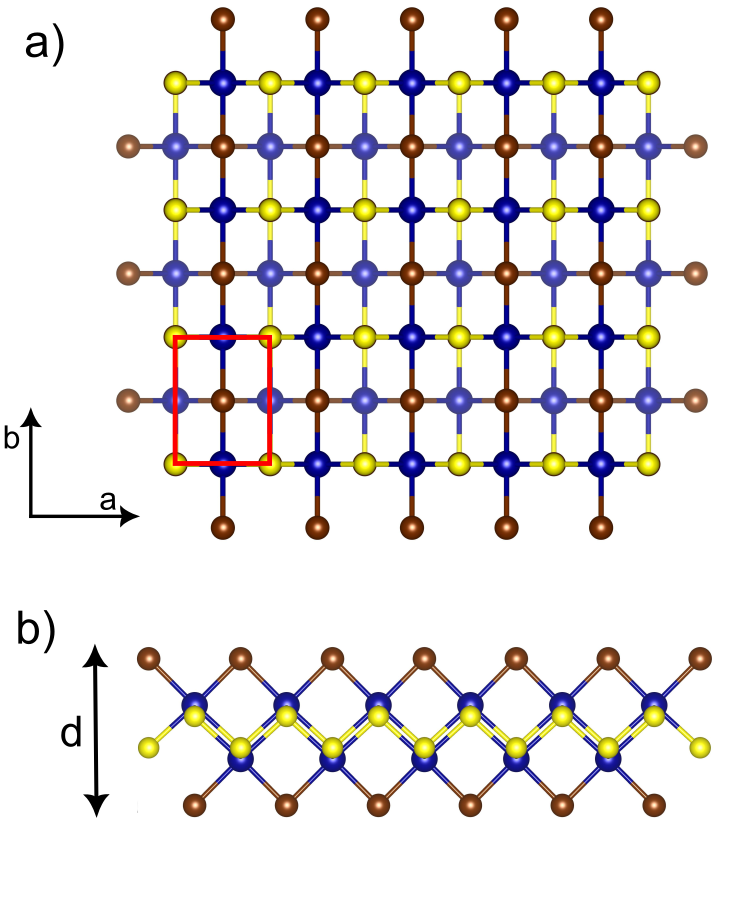}
\caption{(a)~Top and (b)~side views of ML CrSBr; Cr, S, and Br 
         atoms are represented by blue, yellow, and brown spheres, 
         respectively; $d$ indicates the distance between Br planes.}
\label{fig:struct}
\end{figure}

To compute physically meaningful in-plane thermal conductivity, one must 
define an effective thickness of the ML, a choice that is subjected to 
certain degree of arbitrariness. In one-atom thick MLs, such as graphene and 
h-BN, the most widely adopted approach consists in taking as thickness the
interlayer spacing in the bulk. ML CrSBr, on the other hand, is made of 4 
atomic planes and another possible choice is taking the distance between 
the two outer Br planes, $d$ (see Figure~\ref{fig:struct} and Table~\ref{tab:FMvsAFM}).
However, this solution is impractical when it comes to compare the thermal 
conductivity of the ML with the one of few-layer systems or bulk CrSBr.
Therefore, we stick to the usual approach and take the interlayer
separation as the effective ML thickness. To this end, we optimized
the lattice vectors and atomic positions of bulk CrSBr with AA
stacking and took the resulting {\bf c}-vector as the ML thickness.

\section{\label{sec:results}Results}

\begin{table}[t]
\caption{Lattice parameters ($a$ and $b$), ML thickness ($d$) and 
         relative total energy for the FM and AFM configurations 
         of ML CrSBr.}
\label{tab:FMvsAFM}
\begin{tabular}{lcc}
\hline\hline
                         & AFM     & FM    \\
\hline
$\Delta E$ (meV/f.u.)    &  49     & -     \\
$a$ (\AA)                & 3.574  & 3.575  \\
$b$ (\AA)                & 4.817  & 4.806  \\
$d$ (\AA)                & 5.711  & 5.703  \\
\hline
\end{tabular}
\end{table}

\noindent
{\bf Ground-state and response to uniaxial and biaxial stress.} At first we 
optimized the atomic positions and cell vectors of monolayer (ML) CrSBr in 
both the ferromagnetic (FM) and antiferromagnetic (AFM) magnetic ordering, 
finding very similar lattice constants (see Table~\ref{tab:FMvsAFM})
and that the ground-state is FM ($\Delta E_{\rm{FM - AFM}} = -49$~meV/f.u.), 
corroborating previous findings reported in the literature~\cite{GuoNanoscale18,
LeeNL21, RudenkonpjCompMat23}.

We then applied uniaxial tensile and compressive strain ($\epsilon$) up to $\pm 5$\% to the $a$ and $b$ lattice vectors, i.e. the $x$ and $y$-axis, in order to investigate whether the application of an external uniaxial stress can lead to a modulation of the energy difference $\Delta$E$_{\rm FM - AFM}$, and thus induce or ease a magnetic phase transition. In these calculations, the lattice vector perpendicular to the strained one is left free to relax and we found a positive Poisson's ratio, as in conventional non-auxetic materials (i.e., when one lattice vector is compressed, the other expands and viceversa). For each configuration, the total energies of the FM and AFM phases were computed, and the energy difference was monitored as a function of strain. In all cases, the FM state remains the most stable configuration, with $\Delta$E$_{\rm FM - AFM}$ consistently negative across the entire strain range (see Supporting Information, Figure S1).
This demonstrates that, within the investigated range, uniaxial mechanical deformation does not induce any magnetic phase transition in monolayer CrSBr, highlighting the robustness of its ferromagnetic ground state.

\begin{figure}[t]
\includegraphics[width=\linewidth]{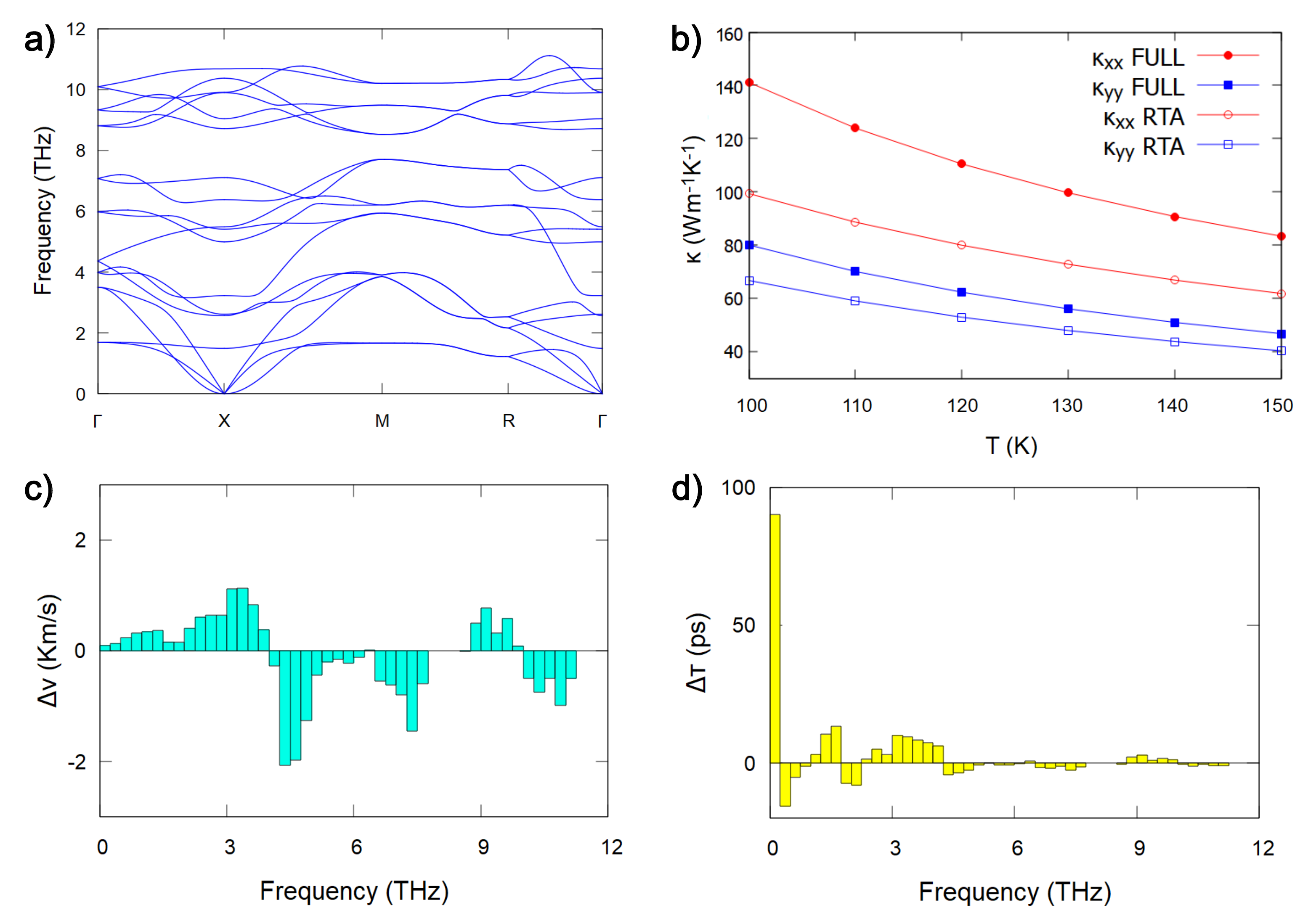}
\caption{(a)~Phonon dispersion. (b)~Thermal conductivity as a function 
         of temperature. (c) $\Delta v = v_x - v_y$ and (d)~$\Delta \tau 
         = \tau_x - \tau_y$ as a function of frequency; before subtraction,
         velocities and lifetimes were averaged over intervals of 1~THz. Phonon lifetimes, 
         $\tau_i$ with $i = x,y$ (see text for definition) are computed at 150~K.}
\label{fig:phonons_kappa_v_tau}
\end{figure}

We further investigated the effect of strain by simultaneously applying it along both the $a$ and $b$ lattice directions to evaluate whether a combined deformation could alter the magnetic ground state of the CrSBr monolayer. Biaxial tensile and compressive strains were again applied in the range of $\pm 5$\%, and the total energies of the FM and AFM configurations were calculated for each strain level. Similar to the uniaxial case, the FM configuration remains energetically favored over the AFM one throughout the entire strain range (Supporting Information, Figure S1).
The energy difference $\Delta$E$_{\rm FM - AFM}$ remains negative, confirming that no magnetic phase transition occurs under biaxial strain either.
These results indicate that the ferromagnetic ground state of monolayer CrSBr is remarkably robust, being insensitive to both uniaxial and biaxial mechanical deformation within the studied strain range.

\noindent
{\bf Phonon dispersion and anisotropic thermal conductivity.} Once the ferromagnetic ground state of the ML was established, we investigated its in-plane thermal conductivity. The phonon dispersion is displayed in Figure~\ref{fig:phonons_kappa_v_tau}a. At variance with previous results, where phonon bands next to the $\Gamma$ point exhibit a small pocket of negative frequencies~\cite{LiuJPCM25, XuanMA23}, by enforcing the rotational sum rules, we obtain the correct quadratic dispersion for the ZA modes, as expected in 2DMs~\cite{CarreteMRL16}.

The thermal conductivity, ${\bm \kappa}$, was computed in the temperature range 100–150 K, where CrSBr is FM (the Curie temperature, above which CrSBr becomes paramagnetic, has been estimated to be 146-150~K~\cite{WangAPL20, LeeNL21, ZhouPCCP25}). At such low temperatures heat transport is dominated by long wavelength phonons and convergence of the lattice thermal conductivity usually requires extremely dense {\bf q}-meshes  to sample the Brillouin zone. We conducted a systematic convergence study (see Supporting Information, Figure S2) of $\kappa_{xx}$ and $\kappa_{yy}$ for samples of different characteristic sizes, $L$ (namely, infinite, 10~$\mu$m, and 100~nm) and reached a robust convergence for a $100 \times 100$ {\bf q}-point grid.
Our results for ${\bm \kappa}$ as a function of temperature are reported in Figure~\ref{fig:phonons_kappa_v_tau}b. Therein, we also compare the predictions of the full solution of the phonon BTE~\cite{FugalloPS18, CarreteVermeerschKatre2017} with those obtained within the simplified RTA approach, finding that the latter provides completely unreliable values: not only the absolute values of $\kappa_{xx}$ and $\kappa_{yy}$, but even the ratio between the two differ significantly from the results of the full solution (the error introduced by the RTA is particularly significative for $\kappa_{xx}$, i.e., $\approx$25\%, almost twice as much the one for $\kappa_{yy}$). 
Indeed, as previously reported, in 2DMs~\cite{Cepellotti2015,FugalloCepellotti2014,Torres_2019} there is a significant fraction of phonon scattering events, involving phonon-phonon interaction, that do not restore first-order deviations from the equilibrium phonon distributions. Normal scattering events fall into this category, but some of the resistive processes do not restore to equilibrium small deviations of the phonon population along the transport direction~\cite{dingZhiwei2018} either. This has two consequences; (a) the RTA underestimates the thermal conductivity by treating some processes as resistive when they are not, as appreciable in Figure~\ref{fig:phonons_kappa_v_tau}b, and (b) since individual phonons no longer diagonalize the scattering matrix, one cannot strictly speak of individual phonon contributions, or cumulative curves. Rather, one must resort to a partially collective description of the system~\cite{torresandbafaluy2017}, or adopt a relaxon picture based on the diagonalization of the full scattering operator ~\cite{CepellottiMarzani2016}.
The fact that our calculations highlight a larger correction to the RTA on $\kappa_{xx}$ is indicative that Normal processes in ML CrSBr would be dominated by phonons that predominantly travel along the $x$-axis.

Our results agree very well with those of Ref.~\citenum{HanJAP25}. 
We found $\kappa_{xx} = 86.31$~W~m$^{-1}$K$^{-1}$ and $\kappa_{yy} 
= 43.08$~W~m$^{-1}$K$^{-1}$ at 150~K, somewhat larger than those of Han 
and coworkers (80.79 and 37.38~W~m$^{-1}$K$^{-1}$ at 145~K), consistent 
with the fact that they took into account also higher-order anharmonic 
scattering processes~\cite{HanJAP25}. The overall good agreement with 
their results, however, suggests that four-phonon processes in ML 
CrSBr are not a significant source of additional scattering.

The thermal conductivity is strongly anisotropic, with an anisotropy ratio $\kappa_{xx}/\kappa_{yy}$ at 150~K of 1.8 (1.5 at the RTA level). In order to see whether the observed anisotropy can be traced back to a difference in phonon velocities, we computed $\Delta v = v_x - v_y$, where $v_i$ is the $i$-component of the group velocity ${\bm v} = \partial \omega/\partial {\bm q}$. Indeed, we find that at low frequency, $\omega \lesssim 4$~THz, $\Delta v > 0$ and thus $v_x > v_y$, in agreement with the larger value of $\kappa_{xx}$ as compared to $\kappa_{yy}$. To factor in the role of anharmonicity via phonon lifetimes, we define $\tau_i = \tau \frac{v_i}{|{\bm v}|}$, where $i =x,y$, so that $\tau_i$ is the lifetime of phonons that predominately propagates along the $i$-axis. Similarly to what we did for the velocities, we computed $\Delta \tau = \tau_x - \tau_y$ and found that at very low frequency $\tau_x >> \tau_y$, which again contributes to determine the ${\bm \kappa}$ anisotropy.

\begin{figure}[t]
\includegraphics[width=\linewidth]{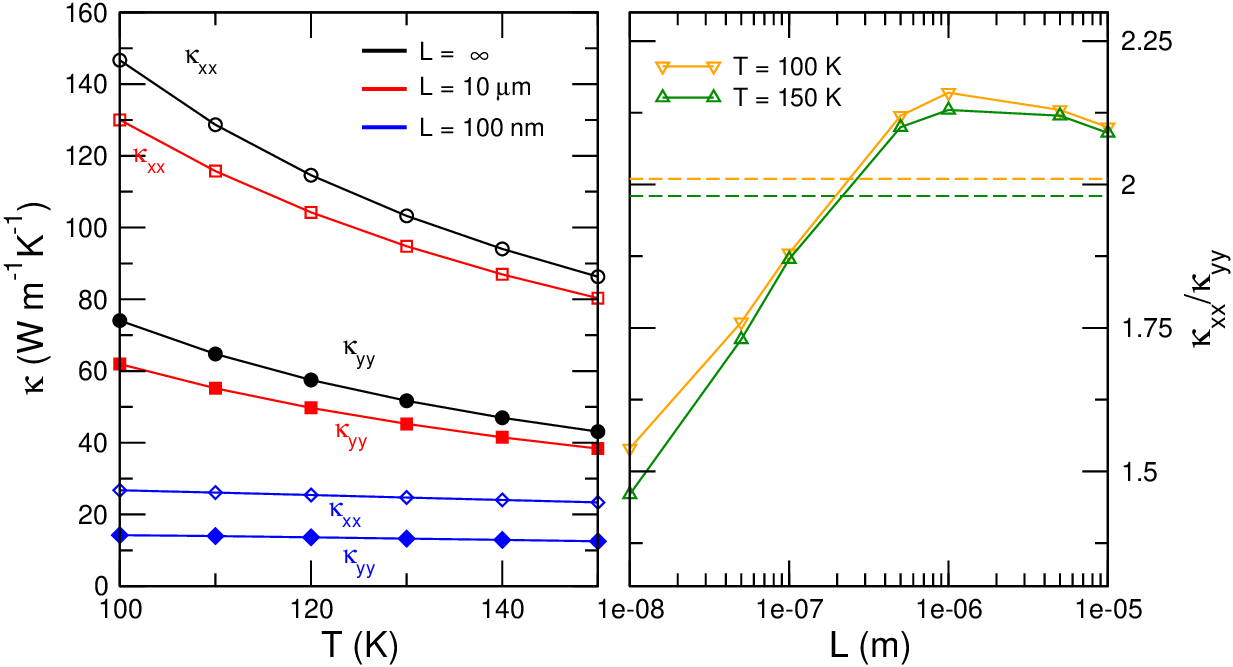}
\caption{(Left)~Thermal conductivity as a function of temperature 
         of ML CrSBr of different characteristic size, $L$. 
         (Right)~Anisotropy ratio as a function of $L$ at $T=100$ 
         and 150~K.}
\label{fig:kappavsT_anisvsL_v2}
\end{figure}

\begin{table}[t]
\begin{center}
\begin{tabular}{lcclcc}
\hline\hline
                          & RTA   & \btef  &  & RTA   & \btef \\
\hline\hline
 $L = \infty$ & & & & & \\
\hline 
 $\kappa^{\rm FM}_{xx}$ & 61.83 & 83.38 & \multirow{2}{*}{$\kappa^{\rm FM}_{xx}/\kappa^{\rm FM}_{yy}$} & \multirow{2}{*}{1.53} & \multirow{2}{*}{1.78} \\
 $\kappa^{\rm FM}_{yy}$ & 40.35  & 46.77  & &&\\
\hline\hline
 $L = 10 \mu$m & & & & & \\
\hline
 $\kappa^{\rm FM}_{xx}$   & 58.22   & 76.57   & \multirow{2}{*}{$\kappa^{\rm FM}_{xx}/\kappa^{\rm FM}_{yy}$} & \multirow{2}{*}{1.62} & \multirow{2}{*}{1.87} \\
 $\kappa^{\rm FM}_{yy}$   & 35.85   & 40.99  & & & \\ 
\hline\hline
 $L = 100 $nm & & & \\
\hline
 $\kappa^{\rm FM}_{xx}$   & 21.01   & 23.09   & \multirow{2}{*}{$\kappa^{\rm FM}_{xx}/\kappa^{\rm FM}_{yy}$} & \multirow{2}{*}{1.67} & \multirow{2}{*}{1.77} \\
 $\kappa^{\rm FM}_{yy}$   & 12.60  & 13.02  & &&\\
\hline
\end{tabular}
\end{center}
\caption{Thermal conductivities at 150~K, in W~m$^{-1}$K$^{-1}$, for the different size configurations. Results are given for the RTA and 
the full solution of the BTE.}
\label{tab:Kappas}
\end{table}

\noindent
{\bf Anisotropy tuning through size modulation.} When studying the thermal conductivity of a 2D material, it is indeed important to consider the dimensions of the flake system, since these directly influence heat transport at the microscopic level. The reason is that the thermal conductivity is significantly influenced by the mean free path (MFP) of phonons. If the flake size is comparable to, or smaller than the MFP of a given set of phonons, the latter will be suppressed, giving rise to a boundary scattering term that must be accounted for. As a matter of fact, ${\bm \kappa}$ converges faster with the number of {\bf q}-points as sample size shrinks down (see Supporting Information, Figure S2), as very long-wavelength phonons progressively cease to contribute to ${\bm \kappa}$ when the sample dimensions are reduced.

Figure~\ref{fig:kappavsT_anisvsL_v2} displays $\kappa_{xx}$ and $\kappa_{yy}$ as a function of temperature of flakes of three selected sizes: $L = \infty$ (the infinite flake, same as in Figure~\ref{fig:phonons_kappa_v_tau}b), $L = 10~\mu$m, and $L =$~100~nm. The effect of boundary scattering in suppressing ${\bm \kappa}$ can be clearly appreciated, with values for the 100~nm flake 7 times smaller than for the unbounded sample. Additionally, we also found that Normal scattering processes are progressively suppressed when considering finite size effects and when $L =$ 100 nm we almost recover the independent phonon picture consubstantial to the RTA (see Table~\ref{tab:Kappas}).

The characteristic flake size, $L$, has also an effect on the anisotropy in thermal transport, as shown in the right panel of Figure~\ref{fig:kappavsT_anisvsL_v2}, where we plot the evolution of $\kappa_{xx}$/$\kappa_{yy}$ as a function of the lateral dimension $L$ of the sample. For small $L$ values ($\sim 10^{-8}$ m), the ratio remains low, indicating a much less anisotropic thermal transport regime. As $L$ increases, $\kappa_{xx}$/$\kappa_{yy}$ progressively rises, staying around 2 and revealing that the intrinsic anisotropy of the material fully develops in the greater surface limit. 

\begin{figure}[t]
\includegraphics[width=\linewidth]{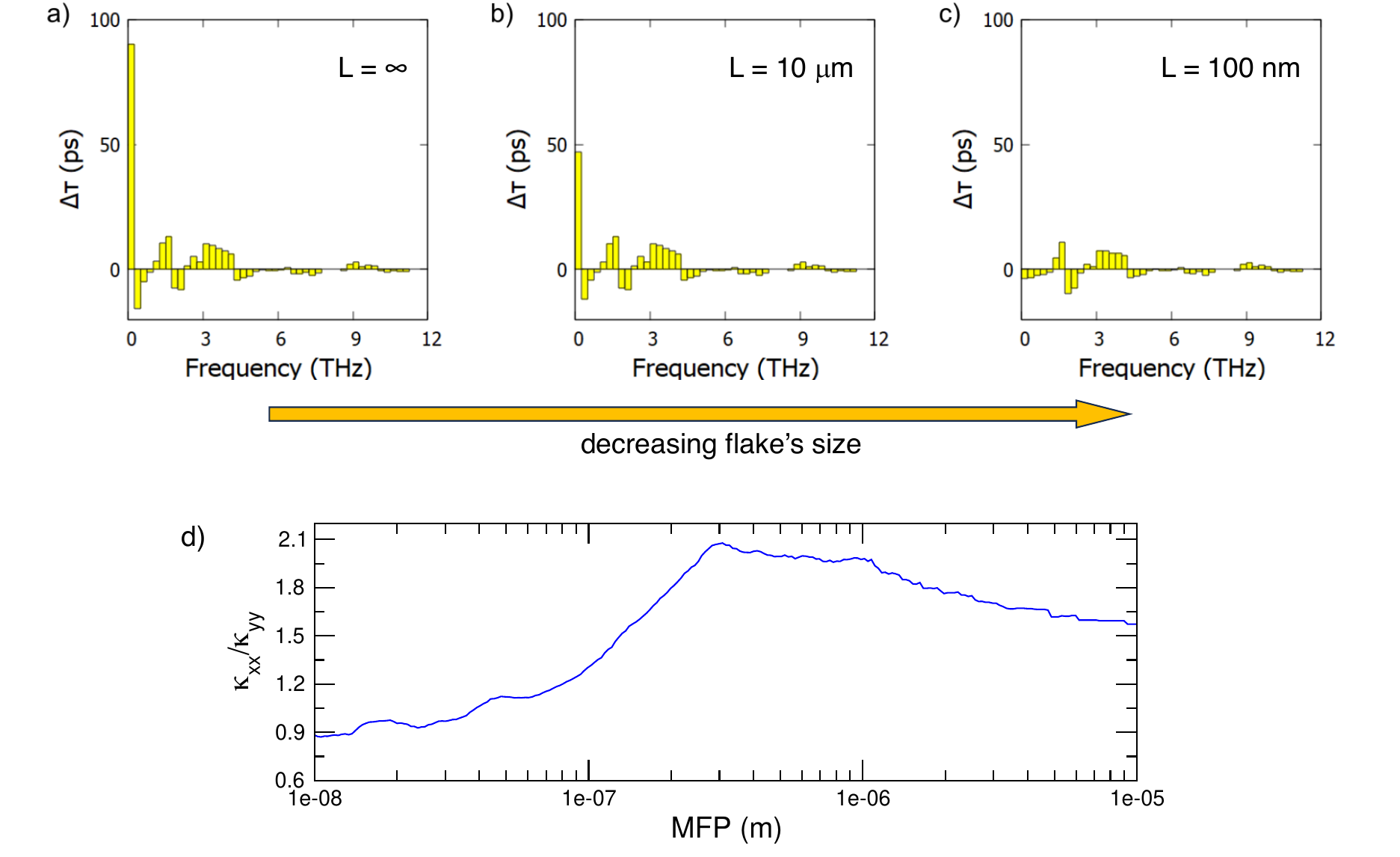}
\caption{$\Delta \tau = \tau_x - \tau_y$ as a function of frequency for 
         (a)~$L = \infty$  (b)~10~$\mu$m, and (c)~100nm; before subtraction,
         lifetimes were averaged over intervals of 1~THz. Phonon lifetimes,
         $\tau_i$ with $i = x,y$ (see text for definition) are computed at 150~K.
         (d)~Ratio between the cumulative thermal conductivity, $\kappa_{xx}$ and 
         $\kappa_{yy}$, as a function of the MFP, $\Lambda$, where $\kappa(\bar{\Lambda})$
         is defined as the thermal conductivity considering only phonons with MFP
         $\Lambda < \bar{\Lambda}$.} 
\label{fig:taus}
\end{figure}

As discussed above, one of the reasons leading to $\kappa_{xx} > \kappa_{yy}$ 
in the infinite flake is that phonons propagating along the $x$-axis have 
longer lifetimes --and therefore larger MFPs-- than those propagating along the 
$y$-axis. Consequently, when the flake size is uniformly reduced, one would
expect boundary scattering to affect $\kappa_{xx}$ sooner than $\kappa_{yy}$, 
since $\kappa_{xx}$ is built up by phonons with longer MFPs. This is indeed
what we observe if we repeat the analysis of Figure~\ref{fig:phonons_kappa_v_tau}d
in the case of smaller samples, e.g. $L = 10 \mu$m and $L = 100$~nm. The results
shown in Figure~\ref{fig:taus}, show that the large low-frequency peak in 
$\Delta \tau$ is progressively quenched as the sample size is reduced and 
that it vanishes at $L = 100$~nm. Notice though, that such a dependence with
the sample size is not fully monotonic and that $\kappa_{xx}/\kappa_{yy}$
exhibits a maximum around $L = 1 \mu$m (see Figure~\ref{fig:kappavsT_anisvsL_v2}). 
This behavior stems from the role played by phonons with MFP between 0.2 and $1 \mu$m
in determining the anisotropy, as shown in Figure~\ref{fig:taus}d, where we plot 
the ratio between the cumulative thermal conductivity, $\kappa_{xx}$ and 
$\kappa_{yy}$, as a function of the phonon MFP.

This behavior demonstrates that the thermal anisotropy can be effectively tuned by controlling the flake’s dimensions. In the nanoscale regime, enhanced phonon–boundary scattering suppresses the in-plane thermal conductivity components differently along $x$ and $y$, thereby reducing the overall anisotropy. Conversely, for larger flakes where boundary effects become negligible, the bulk anisotropy is restored. These results highlight that surface-area modulation offers a viable route to engineer the thermal anisotropy of layered materials, enabling control over heat transport properties via geometric confinement.

\section{\label{sec:concl}Conclusions}

In summary, we have computed the thermal conductivity of the ML CrSBr, 
finding $\kappa_{xx}$ to be almost twice as large as $\kappa_{yy}$ and 
tracing back this strong anisotropy to a combined effects of phonon 
velocities and phonon lifetimes. We also found that, by suppressing 
long MFP phonons, the anisotropy can be tuned to some extent by 
controlling the flake size. Our attempts to stabilize AFM ordering
by means of tensile or compressive uni/biaxial strain were unsuccessful
and ML CrSBr remains FM in all the conditions explored.

\begin{acknowledgement}
We acknowledge financial support by MCIN/AEI/10.13039/501100011033
under grant PDC2023-145934-I00, PID2024-162811NB-I00, PID2024-161603NB-I00, PID2023-146181OB-I00 and the Severo Ochoa Centres of Excellence Program under grant CEX2023-001263-S, and by the Generalitat de Catalunya under grant 2021 SGR 01519. This work is also supported by the Horizon Europe research and innovation program of the European Union under the Marie Sk\l{}odowska-Curie grant agreement 101118915 (TIMES).
Calculations were performed at the Centro de Supercomputaci\'on de Galicia (CESGA).
\end{acknowledgement}

\begin{suppinfo}
The data of unit cell relaxation, the interatomic force constants 
(IFCs) and other input/ouput files of the DFT and BTE calculations
are freely available at dx.doi.org/10.5281/zenodo.18787296.
\end{suppinfo}



\begin{mcitethebibliography}{46}
\providecommand*\natexlab[1]{#1}
\providecommand*\mciteSetBstSublistMode[1]{}
\providecommand*\mciteSetBstMaxWidthForm[2]{}
\providecommand*\mciteBstWouldAddEndPuncttrue
  {\def\EndOfBibitem{\unskip.}}
\providecommand*\mciteBstWouldAddEndPunctfalse
  {\let\EndOfBibitem\relax}
\providecommand*\mciteSetBstMidEndSepPunct[3]{}
\providecommand*\mciteSetBstSublistLabelBeginEnd[3]{}
\providecommand*\EndOfBibitem{}
\mciteSetBstSublistMode{f}
\mciteSetBstMaxWidthForm{subitem}{(\alph{mcitesubitemcount})}
\mciteSetBstSublistLabelBeginEnd
  {\mcitemaxwidthsubitemform\space}
  {\relax}
  {\relax}

\bibitem[Novoselov \latin{et~al.}(2005)Novoselov, Jiang, Schedin, Booth,
  Khotkevich, Morozov, and Geim]{NovoselovJiangSchedin2005}
Novoselov,~K.~S.; Jiang,~D.; Schedin,~F.; Booth,~T.~J.; Khotkevich,~V.~V.;
  Morozov,~S.~V.; Geim,~A.~K. Two-dimensional atomic crystals.
  \emph{Proceedings of the National Academy of Sciences} \textbf{2005},
  \emph{102}, 10451--10453\relax
\mciteBstWouldAddEndPuncttrue
\mciteSetBstMidEndSepPunct{\mcitedefaultmidpunct}
{\mcitedefaultendpunct}{\mcitedefaultseppunct}\relax
\EndOfBibitem
\bibitem[Novoselov \latin{et~al.}(2016)Novoselov, Mishchenko, Carvalho, and
  Neto]{NovoselovScience16}
Novoselov,~K.~S.; Mishchenko,~A.; Carvalho,~A.; Neto,~A. H.~C. 2D materials and
  van der Waals heterostructures. \emph{Science} \textbf{2016}, \emph{353},
  aac9439\relax
\mciteBstWouldAddEndPuncttrue
\mciteSetBstMidEndSepPunct{\mcitedefaultmidpunct}
{\mcitedefaultendpunct}{\mcitedefaultseppunct}\relax
\EndOfBibitem
\bibitem[Novoselov(2011)]{Novoselov2011}
Novoselov,~K.~S. Nobel Lecture: Graphene: Materials in the Flatland. \emph{Rev.
  Mod. Phys.} \textbf{2011}, \emph{83}, 837--849\relax
\mciteBstWouldAddEndPuncttrue
\mciteSetBstMidEndSepPunct{\mcitedefaultmidpunct}
{\mcitedefaultendpunct}{\mcitedefaultseppunct}\relax
\EndOfBibitem
\bibitem[Radisavljevic \latin{et~al.}(2011)Radisavljevic, Radenovic, Brivio,
  Giacometti, and Kis]{RadisavljevicRadenovicBrivio2011}
Radisavljevic,~B.; Radenovic,~A.; Brivio,~J.; Giacometti,~V.; Kis,~A.
  Single-layer MoS2 transistors. \emph{Nature Nanotechnology} \textbf{2011},
  \emph{6}, 147--150\relax
\mciteBstWouldAddEndPuncttrue
\mciteSetBstMidEndSepPunct{\mcitedefaultmidpunct}
{\mcitedefaultendpunct}{\mcitedefaultseppunct}\relax
\EndOfBibitem
\bibitem[Geim and Grigorieva(2013)Geim, and Grigorieva]{GeimGrigorieva2013}
Geim,~A.~K.; Grigorieva,~I.~V. Van der Waals heterostructures. \emph{Nature}
  \textbf{2013}, \emph{499}, 419--425\relax
\mciteBstWouldAddEndPuncttrue
\mciteSetBstMidEndSepPunct{\mcitedefaultmidpunct}
{\mcitedefaultendpunct}{\mcitedefaultseppunct}\relax
\EndOfBibitem
\bibitem[Yang \latin{et~al.}(2022)Yang, Valenzuela, Chshiev, Couet, Dieny,
  Dlubak, Fert, Garello, Jamet, Jeong, Lee, Lee, Martin, Kar, S{\'e}n{\'e}or,
  Shin, and Roche]{Yang2022}
Yang,~H. \latin{et~al.}  Two-dimensional materials prospects for non-volatile
  spintronic memories. \emph{Nature} \textbf{2022}, \emph{606}, 663--673\relax
\mciteBstWouldAddEndPuncttrue
\mciteSetBstMidEndSepPunct{\mcitedefaultmidpunct}
{\mcitedefaultendpunct}{\mcitedefaultseppunct}\relax
\EndOfBibitem
\bibitem[Gibertini \latin{et~al.}(2019)Gibertini, Koperski, Morpurgo, and
  Novoselov]{Gibertini2019}
Gibertini,~M.; Koperski,~M.; Morpurgo,~A.~F.; Novoselov,~K.~S. Magnetic 2D
  materials and heterostructures. \emph{Nature Nanotechnology} \textbf{2019},
  \emph{14}, 408--419\relax
\mciteBstWouldAddEndPuncttrue
\mciteSetBstMidEndSepPunct{\mcitedefaultmidpunct}
{\mcitedefaultendpunct}{\mcitedefaultseppunct}\relax
\EndOfBibitem
\bibitem[Gong \latin{et~al.}(2017)Gong, Li, Li, Ji, Stern, Xia, Cao, Bao, Wang,
  Wang, Qiu, Cava, Louie, Xia, and Zhang]{Gong2017}
Gong,~C.; Li,~L.; Li,~Z.; Ji,~H.; Stern,~A.; Xia,~Y.; Cao,~T.; Bao,~W.;
  Wang,~C.; Wang,~Y.; Qiu,~Z.~Q.; Cava,~R.~J.; Louie,~S.~G.; Xia,~J.; Zhang,~X.
  Discovery of intrinsic ferromagnetism in two-dimensional van der Waals
  crystals. \emph{Nature} \textbf{2017}, \emph{546}, 265--269\relax
\mciteBstWouldAddEndPuncttrue
\mciteSetBstMidEndSepPunct{\mcitedefaultmidpunct}
{\mcitedefaultendpunct}{\mcitedefaultseppunct}\relax
\EndOfBibitem
\bibitem[Huang \latin{et~al.}(2017)Huang, Clark, Navarro-Moratalla, Klein,
  Cheng, Seyler, Zhong, Schmidgall, McGuire, Cobden, Yao, Xiao,
  Jarillo-Herrero, and Xu]{Huang2017}
Huang,~B.; Clark,~G.; Navarro-Moratalla,~E.; Klein,~D.~R.; Cheng,~R.;
  Seyler,~K.~L.; Zhong,~D.; Schmidgall,~E.; McGuire,~M.~A.; Cobden,~D.~H.;
  Yao,~W.; Xiao,~D.; Jarillo-Herrero,~P.; Xu,~X. Layer-dependent ferromagnetism
  in a van der Waals crystal down to the monolayer limit. \emph{Nature}
  \textbf{2017}, \emph{546}, 270--273\relax
\mciteBstWouldAddEndPuncttrue
\mciteSetBstMidEndSepPunct{\mcitedefaultmidpunct}
{\mcitedefaultendpunct}{\mcitedefaultseppunct}\relax
\EndOfBibitem
\bibitem[Wang \latin{et~al.}(2022)Wang, Bedoya-Pinto, Blei, Dismukes, Hamo,
  Jenkins, Koperski, Liu, Sun, Telford, Kim, Augustin, Vool, Yin, Li, Falin,
  Dean, Casanova, Evans, Chshiev, Mishchenko, Petrovic, He, Zhao, Tsen,
  Gerardot, Brotons-Gisbert, Guguchia, Roy, Tongay, Wang, Hasan, Wrachtrup,
  Yacoby, Fert, Parkin, Novoselov, Dai, Balicas, and
  Santos]{WangQingBedoya2022}
Wang,~Q.~H. \latin{et~al.}  The Magnetic Genome of Two-Dimensional van der
  Waals Materials. \emph{ACS Nano} \textbf{2022}, \emph{16}, 6960--7079, PMID:
  35442017\relax
\mciteBstWouldAddEndPuncttrue
\mciteSetBstMidEndSepPunct{\mcitedefaultmidpunct}
{\mcitedefaultendpunct}{\mcitedefaultseppunct}\relax
\EndOfBibitem
\bibitem[Telford \latin{et~al.}(2022)Telford, Dismukes, Dudley, Wiscons, Lee,
  Chica, Ziebel, Han, Yu, Shabani, Scheie, Watanabe, Taniguchi, Xiao, Zhu,
  Pasupathy, Nuckolls, Zhu, Dean, and Roy]{Telford2022}
Telford,~E.~J. \latin{et~al.}  Coupling between magnetic order and charge
  transport in a two-dimensional magnetic semiconductor. \emph{Nature
  Materials} \textbf{2022}, \emph{21}, 754--760\relax
\mciteBstWouldAddEndPuncttrue
\mciteSetBstMidEndSepPunct{\mcitedefaultmidpunct}
{\mcitedefaultendpunct}{\mcitedefaultseppunct}\relax
\EndOfBibitem
\bibitem[Peng \latin{et~al.}(2022)Peng, Lin, Tian, Yang, Zhang, Wang, Gu, Liu,
  Wang, Avdeev, Liu, Zhou, Han, Shen, Yang, Liu, Ye, and Yang]{PengYuxuan2022}
Peng,~Y. \latin{et~al.}  Controlling Spin Orientation and Metamagnetic
  Transitions in Anisotropic van der Waals Antiferromagnet CrPS4 by Hydrostatic
  Pressure. \emph{Advanced Functional Materials} \textbf{2022}, \emph{32},
  2106592\relax
\mciteBstWouldAddEndPuncttrue
\mciteSetBstMidEndSepPunct{\mcitedefaultmidpunct}
{\mcitedefaultendpunct}{\mcitedefaultseppunct}\relax
\EndOfBibitem
\bibitem[Ziebel \latin{et~al.}(2024)Ziebel, Feuer, Cox, Zhu, Dean, and
  Roy]{ZiebelNL24}
Ziebel,~M.~E.; Feuer,~M.~L.; Cox,~J.; Zhu,~X.; Dean,~C.~R.; Roy,~X. CrSBr: An
  Air-Stable, Two-Dimensional Magnetic Semiconductor. \emph{Nano Lett.}
  \textbf{2024}, \emph{24}, 4319--4329\relax
\mciteBstWouldAddEndPuncttrue
\mciteSetBstMidEndSepPunct{\mcitedefaultmidpunct}
{\mcitedefaultendpunct}{\mcitedefaultseppunct}\relax
\EndOfBibitem
\bibitem[Lee \latin{et~al.}(2021)Lee, Dismukes, Telford, Wiscons, Wang, Xu,
  Nuckolls, Dean, Roy, and Zhu]{LeeNL21}
Lee,~K.; Dismukes,~A.~H.; Telford,~E.~J.; Wiscons,~R.~A.; Wang,~J.; Xu,~X.;
  Nuckolls,~C.; Dean,~C.~R.; Roy,~X.; Zhu,~X. Magnetic Order and Symmetry in
  the 2D Semiconductor CrSBr. \emph{Nano Lett.} \textbf{2021}, \emph{21},
  3511--3517\relax
\mciteBstWouldAddEndPuncttrue
\mciteSetBstMidEndSepPunct{\mcitedefaultmidpunct}
{\mcitedefaultendpunct}{\mcitedefaultseppunct}\relax
\EndOfBibitem
\bibitem[Xuan \latin{et~al.}(2023)Xuan, Yang, Du, Zhao, Yan, Liu, Li, and
  Zhang]{XuanMA23}
Xuan,~X.; Yang,~Z.; Du,~R.; Zhao,~Y.; Yan,~Y.; Liu,~C.; Li,~H.; Zhang,~G.
  Ultralow thermal conductivity and anharmonic rattling in two-dimensional CrSX
  (X = Cl{,} Br{,} I) monolayers. \emph{Mater. Adv.} \textbf{2023}, \emph{4},
  4852--4859\relax
\mciteBstWouldAddEndPuncttrue
\mciteSetBstMidEndSepPunct{\mcitedefaultmidpunct}
{\mcitedefaultendpunct}{\mcitedefaultseppunct}\relax
\EndOfBibitem
\bibitem[Liu \latin{et~al.}(2025)Liu, Zhi, Liu, Liu, Jiang, and
  Zhao]{LiuJPCM25}
Liu,~Y.; Zhi,~Y.; Liu,~Q.; Liu,~Y.; Jiang,~X.; Zhao,~J. Lattice thermal
  conductivity in CrSBr: the effects of interlayer interaction, magnetic
  ordering and external strain. \emph{J. Phys.: Condens. Matter} \textbf{2025},
  \emph{37}, 125701\relax
\mciteBstWouldAddEndPuncttrue
\mciteSetBstMidEndSepPunct{\mcitedefaultmidpunct}
{\mcitedefaultendpunct}{\mcitedefaultseppunct}\relax
\EndOfBibitem
\bibitem[Li \latin{et~al.}(2014)Li, Carrete, Katcho, and Mingo]{LiCPC14}
Li,~W.; Carrete,~J.; Katcho,~N.~A.; Mingo,~N. {ShengBTE:} a solver of the
  {B}oltzmann transport equation for phonons. \emph{Comp. Phys. Commun.}
  \textbf{2014}, \emph{185}, 1747--1758\relax
\mciteBstWouldAddEndPuncttrue
\mciteSetBstMidEndSepPunct{\mcitedefaultmidpunct}
{\mcitedefaultendpunct}{\mcitedefaultseppunct}\relax
\EndOfBibitem
\bibitem[Togo \latin{et~al.}(2023)Togo, Chaput, Tadano, and Tanaka]{TogoJPCM23}
Togo,~A.; Chaput,~L.; Tadano,~T.; Tanaka,~I. Implementation strategies in
  phonopy and phono3py. \emph{J. Phys. Condens. Matter} \textbf{2023},
  \emph{35}, 353001\relax
\mciteBstWouldAddEndPuncttrue
\mciteSetBstMidEndSepPunct{\mcitedefaultmidpunct}
{\mcitedefaultendpunct}{\mcitedefaultseppunct}\relax
\EndOfBibitem
\bibitem[McGaughey \latin{et~al.}(2025)McGaughey, Lindsay, Bao, Hamakawa,
  Juneja, Li, Li, Masuki, Meng, Meng, Pandey, Shao, Shiomi, Tadano, Togo, Wang,
  and Zhang]{McGaugheyJAP25}
McGaughey,~A. J.~H. \latin{et~al.}  Phonon Olympics: Phonon property and
  lattice thermal conductivity benchmarking from open-source packages. \emph{J.
  Appl. Phys.} \textbf{2025}, \emph{138}, 135108\relax
\mciteBstWouldAddEndPuncttrue
\mciteSetBstMidEndSepPunct{\mcitedefaultmidpunct}
{\mcitedefaultendpunct}{\mcitedefaultseppunct}\relax
\EndOfBibitem
\bibitem[Han \latin{et~al.}(2025)Han, Li, Tang, Wang, Li, He, Tang, and
  Ouyang]{HanJAP25}
Han,~L.; Li,~Z.; Tang,~Z.; Wang,~X.; Li,~J.; He,~C.; Tang,~C.; Ouyang,~T.
  Notable impact of magnetic order and flat phonon mode on the thermal
  transport properties of 2D magnetic semiconductor CrSBr. \emph{J. Appl.
  Phys.} \textbf{2025}, \emph{138}, 195101\relax
\mciteBstWouldAddEndPuncttrue
\mciteSetBstMidEndSepPunct{\mcitedefaultmidpunct}
{\mcitedefaultendpunct}{\mcitedefaultseppunct}\relax
\EndOfBibitem
\bibitem[Hohenberg and Kohn(1964)Hohenberg, and Kohn]{HohenbergKohn1964}
Hohenberg,~P.; Kohn,~W. Inhomogeneous Electron Gas. \emph{Phys. Rev.}
  \textbf{1964}, \emph{136}, B864--B871\relax
\mciteBstWouldAddEndPuncttrue
\mciteSetBstMidEndSepPunct{\mcitedefaultmidpunct}
{\mcitedefaultendpunct}{\mcitedefaultseppunct}\relax
\EndOfBibitem
\bibitem[Kohn and Sham(1965)Kohn, and Sham]{KohnSham1965}
Kohn,~W.; Sham,~L.~J. Self-Consistent Equations Including Exchange and
  Correlation Effects. \emph{Phys. Rev.} \textbf{1965}, \emph{140},
  A1133--A1138\relax
\mciteBstWouldAddEndPuncttrue
\mciteSetBstMidEndSepPunct{\mcitedefaultmidpunct}
{\mcitedefaultendpunct}{\mcitedefaultseppunct}\relax
\EndOfBibitem
\bibitem[Wang \latin{et~al.}(2021)Wang, Xu, Liu, Tang, and Geng]{VASPKIT}
Wang,~V.; Xu,~N.; Liu,~J.-C.; Tang,~G.; Geng,~W.-T. VASPKIT: A user-friendly
  interface facilitating high-throughput computing and analysis using VASP
  code. \emph{Computer Physics Communications} \textbf{2021}, \emph{267},
  108033\relax
\mciteBstWouldAddEndPuncttrue
\mciteSetBstMidEndSepPunct{\mcitedefaultmidpunct}
{\mcitedefaultendpunct}{\mcitedefaultseppunct}\relax
\EndOfBibitem
\bibitem[Perdew \latin{et~al.}(1996)Perdew, Burke, and
  Ernzerhof]{PerdewBurkeErnzerhof1996}
Perdew,~J.~P.; Burke,~K.; Ernzerhof,~M. Generalized Gradient Approximation Made
  Simple. \emph{Phys. Rev. Lett.} \textbf{1996}, \emph{77}, 3865--3868\relax
\mciteBstWouldAddEndPuncttrue
\mciteSetBstMidEndSepPunct{\mcitedefaultmidpunct}
{\mcitedefaultendpunct}{\mcitedefaultseppunct}\relax
\EndOfBibitem
\bibitem[Grimme \latin{et~al.}(2010)Grimme, Antony, Ehrlich, and
  Krieg]{GrimmeStefanAntony}
Grimme,~S.; Antony,~J.; Ehrlich,~S.; Krieg,~H. A consistent and accurate ab
  initio parametrization of density functional dispersion correction (DFT-D)
  for the 94 elements H-Pu. \emph{The Journal of Chemical Physics}
  \textbf{2010}, \emph{132}, 154104\relax
\mciteBstWouldAddEndPuncttrue
\mciteSetBstMidEndSepPunct{\mcitedefaultmidpunct}
{\mcitedefaultendpunct}{\mcitedefaultseppunct}\relax
\EndOfBibitem
\bibitem[Dudarev \latin{et~al.}(1998)Dudarev, Botton, Savrasov, Humphreys, and
  Sutton]{Dudarev_at_all}
Dudarev,~S.~L.; Botton,~G.~A.; Savrasov,~S.~Y.; Humphreys,~C.~J.; Sutton,~A.~P.
  Electron-energy-loss spectra and the structural stability of nickel oxide: An
  LSDA+U study. \emph{Phys. Rev. B} \textbf{1998}, \emph{57}, 1505--1509\relax
\mciteBstWouldAddEndPuncttrue
\mciteSetBstMidEndSepPunct{\mcitedefaultmidpunct}
{\mcitedefaultendpunct}{\mcitedefaultseppunct}\relax
\EndOfBibitem
\bibitem[Yang \latin{et~al.}(2021)Yang, Wang, Liu, Lu, and Wu]{YangPRB21}
Yang,~K.; Wang,~G.; Liu,~L.; Lu,~D.; Wu,~H. Triaxial magnetic anisotropy in the
  two-dimensional ferromagnetic semiconductor CrSBr. \emph{Phys. Rev. B}
  \textbf{2021}, \emph{104}, 144416\relax
\mciteBstWouldAddEndPuncttrue
\mciteSetBstMidEndSepPunct{\mcitedefaultmidpunct}
{\mcitedefaultendpunct}{\mcitedefaultseppunct}\relax
\EndOfBibitem
\bibitem[Carrete \latin{et~al.}(2017)Carrete, Vermeersch, Katre, {van
  Roekeghem}, Wang, Madsen, and Mingo]{CarreteVermeerschKatre2017}
Carrete,~J.; Vermeersch,~B.; Katre,~A.; {van Roekeghem},~A.; Wang,~T.;
  Madsen,~G.~K.; Mingo,~N. almaBTE : A solver of the space–time dependent
  Boltzmann transport equation for phonons in structured materials.
  \emph{Computer Physics Communications} \textbf{2017}, \emph{220},
  351--362\relax
\mciteBstWouldAddEndPuncttrue
\mciteSetBstMidEndSepPunct{\mcitedefaultmidpunct}
{\mcitedefaultendpunct}{\mcitedefaultseppunct}\relax
\EndOfBibitem
\bibitem[Togo(2023)]{phonopy2-phono3py-JPSJ}
Togo,~A. First-principles Phonon Calculations with Phonopy and Phono3py.
  \emph{J. Phys. Soc. Jpn.} \textbf{2023}, \emph{92}, 012001\relax
\mciteBstWouldAddEndPuncttrue
\mciteSetBstMidEndSepPunct{\mcitedefaultmidpunct}
{\mcitedefaultendpunct}{\mcitedefaultseppunct}\relax
\EndOfBibitem
\bibitem[Eriksson \latin{et~al.}(2019)Eriksson, Fransson, and
  Erhart]{ErikssonATS19}
Eriksson,~F.; Fransson,~E.; Erhart,~P. The Hiphive Package for the Extraction
  of High-Order Force Constants by Machine Learning. \emph{Adv. Theory Simul.}
  \textbf{2019}, \emph{2}, 1800184\relax
\mciteBstWouldAddEndPuncttrue
\mciteSetBstMidEndSepPunct{\mcitedefaultmidpunct}
{\mcitedefaultendpunct}{\mcitedefaultseppunct}\relax
\EndOfBibitem
\bibitem[Carrete \latin{et~al.}(2016)Carrete, Li, Lindsay, Broido, Gallego, and
  Mingo]{CarreteMRL16}
Carrete,~J.; Li,~W.; Lindsay,~L.; Broido,~D.~A.; Gallego,~L.~J.; Mingo,~N.
  Physically founded phonon dispersions of few-layer materials and the case of
  borophene. \emph{Mater. Res. Lett.} \textbf{2016}, \emph{4}, 204--211\relax
\mciteBstWouldAddEndPuncttrue
\mciteSetBstMidEndSepPunct{\mcitedefaultmidpunct}
{\mcitedefaultendpunct}{\mcitedefaultseppunct}\relax
\EndOfBibitem
\bibitem[S\'anchez-Portal and Hern\'andez(2002)S\'anchez-Portal, and
  Hern\'andez]{SanchezPortalPRB02}
S\'anchez-Portal,~D.; Hern\'andez,~E. Vibrational properties of single-wall
  nanotubes and monolayers of hexagonal BN. \emph{Phys. Rev. B} \textbf{2002},
  \emph{66}, 235415\relax
\mciteBstWouldAddEndPuncttrue
\mciteSetBstMidEndSepPunct{\mcitedefaultmidpunct}
{\mcitedefaultendpunct}{\mcitedefaultseppunct}\relax
\EndOfBibitem
\bibitem[Sohier \latin{et~al.}(2017)Sohier, Gibertini, Calandra, Mauri, and
  Marzari]{SohierNL17}
Sohier,~T.; Gibertini,~M.; Calandra,~M.; Mauri,~F.; Marzari,~N. Breakdown of
  Optical Phonons' Splitting in Two-Dimensional Materials. \emph{Nano Lett.}
  \textbf{2017}, \emph{17}, 3758--3763\relax
\mciteBstWouldAddEndPuncttrue
\mciteSetBstMidEndSepPunct{\mcitedefaultmidpunct}
{\mcitedefaultendpunct}{\mcitedefaultseppunct}\relax
\EndOfBibitem
\bibitem[De~Luca \latin{et~al.}(2020)De~Luca, Cartoix{\`{a}}, Indolese,
  Mart{\'{\i}}n-S{\'{a}}nchez, Watanabe, Taniguchi, Schönenberger, Trotta,
  Rurali, and Zardo]{DeLuca2DMater20}
De~Luca,~M.; Cartoix{\`{a}},~X.; Indolese,~D.~I.;
  Mart{\'{\i}}n-S{\'{a}}nchez,~J.; Watanabe,~K.; Taniguchi,~T.;
  Schönenberger,~C.; Trotta,~R.; Rurali,~R.; Zardo,~I. Experimental
  demonstration of the suppression of optical phonon splitting in 2D materials
  by Raman spectroscopy. \emph{2D Mater.} \textbf{2020}, \emph{7}, 035017\relax
\mciteBstWouldAddEndPuncttrue
\mciteSetBstMidEndSepPunct{\mcitedefaultmidpunct}
{\mcitedefaultendpunct}{\mcitedefaultseppunct}\relax
\EndOfBibitem
\bibitem[Guo \latin{et~al.}(2018)Guo, Zhang, Yuan, Wang, and
  Wang]{GuoNanoscale18}
Guo,~Y.; Zhang,~Y.; Yuan,~S.; Wang,~B.; Wang,~J. Chromium sulfide halide
  monolayers: intrinsic ferromagnetic semiconductors with large spin
  polarization and high carrier mobility. \emph{Nanoscale} \textbf{2018},
  \emph{10}, 18036--18042\relax
\mciteBstWouldAddEndPuncttrue
\mciteSetBstMidEndSepPunct{\mcitedefaultmidpunct}
{\mcitedefaultendpunct}{\mcitedefaultseppunct}\relax
\EndOfBibitem
\bibitem[Rudenko \latin{et~al.}(2023)Rudenko, R\"osner, and
  Katsnelson]{RudenkonpjCompMat23}
Rudenko,~A.~N.; R\"osner,~M.; Katsnelson,~M.~I. Dielectric tunability of
  magnetic properties in orthorhombic ferromagnetic monolayer CrSBr. \emph{npj
  Comput. Mater.} \textbf{2023}, \emph{9}, 83\relax
\mciteBstWouldAddEndPuncttrue
\mciteSetBstMidEndSepPunct{\mcitedefaultmidpunct}
{\mcitedefaultendpunct}{\mcitedefaultseppunct}\relax
\EndOfBibitem
\bibitem[Wang \latin{et~al.}(2020)Wang, Qi, and Qian]{WangAPL20}
Wang,~H.; Qi,~J.; Qian,~X. Electrically tunable high Curie temperature
  two-dimensional ferromagnetism in van der Waals layered crystals. \emph{Appl.
  Phys. Lett.} \textbf{2020}, \emph{117}, 083102\relax
\mciteBstWouldAddEndPuncttrue
\mciteSetBstMidEndSepPunct{\mcitedefaultmidpunct}
{\mcitedefaultendpunct}{\mcitedefaultseppunct}\relax
\EndOfBibitem
\bibitem[Zhou and Zhou(2025)Zhou, and Zhou]{ZhouPCCP25}
Zhou,~X.; Zhou,~B. Enhanced Curie temperature of ferromagnetic CrSBr by
  interfacial coupling with elemental two-dimensional ferroelectrics:
  triggering a new p-d super-exchange coupling path. \emph{Phys. Chem. Chem.
  Phys.} \textbf{2025}, \emph{27}, 20500--20508\relax
\mciteBstWouldAddEndPuncttrue
\mciteSetBstMidEndSepPunct{\mcitedefaultmidpunct}
{\mcitedefaultendpunct}{\mcitedefaultseppunct}\relax
\EndOfBibitem
\bibitem[Fugallo and Colombo(2018)Fugallo, and Colombo]{FugalloPS18}
Fugallo,~G.; Colombo,~L. Calculating lattice thermal conductivity: a synopsis.
  \emph{Phys. Scr.} \textbf{2018}, \emph{93}, 043002\relax
\mciteBstWouldAddEndPuncttrue
\mciteSetBstMidEndSepPunct{\mcitedefaultmidpunct}
{\mcitedefaultendpunct}{\mcitedefaultseppunct}\relax
\EndOfBibitem
\bibitem[Cepellotti \latin{et~al.}(2015)Cepellotti, Fugallo, Paulatto, Lazzeri,
  Mauri, and Marzari]{Cepellotti2015}
Cepellotti,~A.; Fugallo,~G.; Paulatto,~L.; Lazzeri,~M.; Mauri,~F.; Marzari,~N.
  Phonon hydrodynamics in two-dimensional materials. \emph{Nature
  Communications} \textbf{2015}, \emph{6}, 6400\relax
\mciteBstWouldAddEndPuncttrue
\mciteSetBstMidEndSepPunct{\mcitedefaultmidpunct}
{\mcitedefaultendpunct}{\mcitedefaultseppunct}\relax
\EndOfBibitem
\bibitem[Fugallo \latin{et~al.}(2014)Fugallo, Cepellotti, Paulatto, Lazzeri,
  Marzari, and Mauri]{FugalloCepellotti2014}
Fugallo,~G.; Cepellotti,~A.; Paulatto,~L.; Lazzeri,~M.; Marzari,~N.; Mauri,~F.
  Thermal Conductivity of Graphene and Graphite: Collective Excitations and
  Mean Free Paths. \emph{Nano Letters} \textbf{2014}, \emph{14}, 6109--6114,
  PMID: 25343716\relax
\mciteBstWouldAddEndPuncttrue
\mciteSetBstMidEndSepPunct{\mcitedefaultmidpunct}
{\mcitedefaultendpunct}{\mcitedefaultseppunct}\relax
\EndOfBibitem
\bibitem[Torres \latin{et~al.}(2019)Torres, Alvarez, Cartoixà, and
  Rurali]{Torres_2019}
Torres,~P.; Alvarez,~F.~X.; Cartoixà,~X.; Rurali,~R. Thermal conductivity and
  phonon hydrodynamics in transition metal dichalcogenides from
  first-principles. \emph{2D Materials} \textbf{2019}, \emph{6}, 035002\relax
\mciteBstWouldAddEndPuncttrue
\mciteSetBstMidEndSepPunct{\mcitedefaultmidpunct}
{\mcitedefaultendpunct}{\mcitedefaultseppunct}\relax
\EndOfBibitem
\bibitem[Ding \latin{et~al.}(2018)Ding, Zhou, Song, Li, Liu, and
  Chen]{dingZhiwei2018}
Ding,~Z.; Zhou,~J.; Song,~B.; Li,~M.; Liu,~T.-H.; Chen,~G. Umklapp scattering
  is not necessarily resistive. \emph{Phys. Rev. B} \textbf{2018}, \emph{98},
  180302\relax
\mciteBstWouldAddEndPuncttrue
\mciteSetBstMidEndSepPunct{\mcitedefaultmidpunct}
{\mcitedefaultendpunct}{\mcitedefaultseppunct}\relax
\EndOfBibitem
\bibitem[Torres \latin{et~al.}(2017)Torres, Torell\'o, Bafaluy, Camacho,
  Cartoix\`a, and Alvarez]{torresandbafaluy2017}
Torres,~P.; Torell\'o,~A.; Bafaluy,~J.; Camacho,~J.; Cartoix\`a,~X.;
  Alvarez,~F.~X. First principles kinetic-collective thermal conductivity of
  semiconductors. \emph{Phys. Rev. B} \textbf{2017}, \emph{95}, 165407\relax
\mciteBstWouldAddEndPuncttrue
\mciteSetBstMidEndSepPunct{\mcitedefaultmidpunct}
{\mcitedefaultendpunct}{\mcitedefaultseppunct}\relax
\EndOfBibitem
\bibitem[Cepellotti and Marzari(2016)Cepellotti, and
  Marzari]{CepellottiMarzani2016}
Cepellotti,~A.; Marzari,~N. Thermal Transport in Crystals as a Kinetic Theory
  of Relaxons. \emph{Phys. Rev. X} \textbf{2016}, \emph{6}, 041013\relax
\mciteBstWouldAddEndPuncttrue
\mciteSetBstMidEndSepPunct{\mcitedefaultmidpunct}
{\mcitedefaultendpunct}{\mcitedefaultseppunct}\relax
\EndOfBibitem
\end{mcitethebibliography}

\providecommand{\latin}[1]{#1}
\makeatletter
\providecommand{\doi}
  {\begingroup\let\do\@makeother\dospecials
  \catcode`\{=1 \catcode`\}=2 \doi@aux}
\providecommand{\doi@aux}[1]{\endgroup\texttt{#1}}
\makeatother
\providecommand*\mcitethebibliography{\thebibliography}
\csname @ifundefined\endcsname{endmcitethebibliography}
  {\let\endmcitethebibliography\endthebibliography}{}

\end{document}